\documentclass[conference,twocolumn]{IEEEtran}

\usepackage[letterpaper, left=1in, right=1in, bottom=1in, top=0.75in]{geometry}

\usepackage{mathrsfs}
\usepackage{mathtools}
\usepackage{amsmath}
\usepackage{amsthm}
\usepackage{amssymb}
\usepackage{graphicx}

\usepackage{cite}

\usepackage{amsmath,amssymb,amsfonts}
\usepackage{algorithm}
\usepackage{algorithmic}

\usepackage{graphicx}
\usepackage{textcomp}
\usepackage{xcolor}

\makeatletter

\providecommand{\tabularnewline}{\\}

\theoremstyle{plain}
\newtheorem{thm}{\protect\theoremname}

\theoremstyle{plain}
\newtheorem{prop}[thm]{Proposition}

\usepackage[caption=false,font=footnotesize]{subfig}

\makeatother

\providecommand{\theoremname}{Theorem}

\begin{document}
\title{Decision Set Optimization and Energy-Efficient MIMO Communications}
\author{\IEEEauthorblockN{Hang~Zou\IEEEauthorrefmark{1}, Chao~Zhang\IEEEauthorrefmark{1},
Samson~Lasaulce\IEEEauthorrefmark{1}, Lucas~Saludjian\IEEEauthorrefmark{2}
and Patrick~Panciatici\IEEEauthorrefmark{2}}\IEEEauthorblockA{\IEEEauthorrefmark{1}LSS, CNRS-CentraleSupelec-Univ. Paris Sud, Gif-sur-Yvette,
France\\
Email: \{hang.zou chao.zhang samson.lasaulce\}@l2s.centralesupelec.fr}\IEEEauthorblockA{\IEEEauthorrefmark{2}RTE, France\\
}}
\maketitle

\begin{abstract}
Assuming that the number of possible decisions for a transmitter (e.g., the number of possible beamforming vectors) has to be finite and is given, this paper investigates for the first time the problem of determining the best decision set when energy-efficiency maximization is pursued. We propose a framework to find a good (finite) decision set which induces a minimal performance loss w.r.t. to the continuous case. We exploit this framework for a scenario of energy-efficient MIMO communications in which transmit power and beamforming vectors have to be adapted jointly to the channel given under finite-rate feedback. To determine a good decision set we propose an algorithm which combines the approach of Invasive Weed Optimization (IWO) and an Evolutionary Algorithm (EA). We provide a numerical analysis which illustrates the benefits of our point of view. In particular, given a performance loss level, the feedback rate can by reduced by $2$ when the transmit decision set has been designed properly by using our algorithm. The impact on energy-efficiency is also seen to be significant. 
\end{abstract}


The literature of wireless communications comprises a large number of works on resource allocation problems at the transmitter side. The transmitter may have to choose e.g., its transmission power, its precoding matrix, or its modulation coding scheme (MCS). Almost always the corresponding decision set is taken for granted, as a data of the problem. In the present work, we show this might be questionable and that the choice of optimization set itself may be optimized. In fact, many practical reasons for operating with a finite decision set at the transmitter may be given: computational complexity limitations at the transmitter, finite feedback-rate limitations, implementability requirements, robustness needs, standardization requirements, etc. The motivation retained in this paper is the presence of a finite-rate feedback (FRF) mechanism. One of reasons for this is that the design of feedback mechanisms, protocols, or channels attracts a lot of attention of the community. The benefit from designing carefully the feedback system is of particular interest for 5G communications for which high data rates are pursued while managing very efficiently the available energy resources. 

When inspecting the literature of feedback mechanisms, it is seen that some successful analog feedback mechanisms have been proposed (see e.g.,\cite{caire-FACSSC-2006, Marzetta-TSP-2006}) but more and more attention has been given to digital feedback techniques, a.k.a., limited-rate feedback (LRF) or FRF \cite{Roh-TIT-2006,jin-TIT-2006}. Although the problem of resource allocation at the transmitter under FRF is quite well-known, the problem of designing the transmitter decision set under FRF has been left largely unexplored. The closest works to the one reported in this paper seem to be those concerning the maximization of capacity of MIMO communications. In particular, in \cite{love-TIT-2003}\cite{jin-TIT-2006}, the authors study the problem of designing the transmit codebook in presence of FRF. In \cite{love-TIT-2003}, it is shown that the best beamforming vectors can be obtained over a Grassmannian manifold and the capacity loss induced by limited feedback can be upper bounded in a tight manner. In \cite{jin-TIT-2006}, a fundamental performance analysis is conducted to assess the loss in terms of sum-capacity of a MIMO broadcast channel in presence of FRF.  A new Lloyd-Max-type algorithm is proposed by treating the beamforming vector selection problem as a vector quantization (VQ) problem for capacity maximization under FRF in \cite{Roh-TIT-2006}. The ergodic secrecy capacity of a wiretap channel is investigated under FRF in \cite{hyadi-TWC-2017}, as an application to secure communications. 

Although other works on the design of the codebook for capacity optimization might be cited,  few papers (see \cite{amin-CL-2014}) have addressed the problem of designing a codebook for channel state information (CSI) feedback to have an energy-efficient MIMO communication. It appears that almost all papers of the authors' knowledge assume a given set or space for the possible transmit vectors used to maximize energy-efficiency \cite{zapp-TSP-2017}. Motivated by this observation, one of the objectives of this paper is to propose a methodology to determine a good decision set which can be applied to maximize MIMO energy-efficiency but also to maximize more general utility functions. 

The rest of the paper is organized as follows.  The system model and the  problem of finding a good decision set for energy-efficient MIMO communications is introduced in Sec.~\ref{sec:sys_problem}. To effectively determine the optimal decision set for this scenario, we propose in Sec.~\ref{sec:IWO-DE-algorithm} an algorithm which combines the approach of Invasive Weed Optimization (IWO) and an evolutionary algorithm; this algorithm may be applied to more general utility functions. In Sec.~\ref{sec:Simulation-Results}, we provide a numerical perform analysis which strongly supports our framework and shows the benefits e.g., in terms of feedback rate from designing the transmission decision set carefully.

%The performance is measured in terms of a utility function $u(x;g)$, $x$ (resp. $g$)  being the  decision (resp. the parameter) when $g$ is acquired through a finite-rate channel
%
%It is well known that even a small amount of feedback information can significantly improve the performance of the system, e.g.,  in terms of capacity, success probability, energy-efficiency (EE) in \cite{love-TIT-2003,jin-TIT-2006,love-TOC-2003,amin-CL-2014}. The benefit from designing carefully the feedback system is of particular interest for 5G communications for which high data rates are pursued while managing very efficiently the available energy resources. 

%feedback system 
%role of feedback then list different utility function with different feedback and their achievement.
%Lastly point out that there is no general mechanism for feedback system design especially for what should be feedback and how to feedback.

%5g importance of EE, and no more interesting results in the feedback design of in the sense of EE maximization.

\section{Problem formulation\label{sec:sys_problem}}

The considered communication scenario comprises a multi-antenna transmitter which has to adapt the transmit power $p \in [0, P_{\max}]$ and its unit beamforming vector $\boldsymbol{\boldsymbol{\omega}}\in\mathbb{C}^{N_{t}\times1}$ ($\left\Vert \boldsymbol{\boldsymbol{\omega}}\right\Vert  = 1$) to the realization of the channel transfer matrix $\mathbf{H}\in \mathbb{C}^{N_{r}\times N_{t}}$, $N_{t}$ and $N_{r}$ being respectively the number of transmit antennas and receive antennas. We assume that every entry of $\mathbf{H}$ is circularly symmetric complex Gaussian distributed according to $\mathcal{CN}\left(0,1\right)$. The \textit{action or decision} of the transmitter is thus given by the pair $x=\left(p,\boldsymbol{\boldsymbol{\boldsymbol{\omega}}}\right)$. The objective of the transmitter is to maximize its energy-efficiency by adapting its decision to the channel. A very common measure of energy-efficiency is given by the ratio of a benefit function (e.g., the packet success rate or a measure of the transmission rate) to a cost power (e.g., an increasing function of the radiated power). The assumed utility function has the following form:
\begin{equation}
u^{i}\left(x;  \mathbf{H} \right)\coloneqq\frac{V^{i}\left(x; \mathbf{H} \right)}{C\left(x\right)}\label{eq:general form of EE}
\end{equation}
where $V^{i}(x;\mathbf{H})$ is the transmission benefit obtained from choosing decision $x$ over a channel matrix $\mathbf{H}$ and $C\left(x\right)$ the transmission cost of using decision $x$; $i$ stands for the considered case index, the two cases being defined just next. Indeed, for the \textit{benefit function}, we will use one of the following functions: 
\begin{itemize}
\item \textit{Case I benefit function (capacity function):} $V^{\textrm{I}}\left(p,\boldsymbol{\boldsymbol{\boldsymbol{\omega}}};\mathbf{H}\right)=\log\left(1+\frac{p\left\Vert \mathbf{H}\boldsymbol{\boldsymbol{\boldsymbol{\omega}}}\right\Vert ^{2}}{\sigma^{2}}\right)$ (see e.g., \cite{vero-TSP-2011}\cite{zapp-TSP-2017}). 
\item \textit{Case II benefit function (packet success rate)}: $V^{\textrm{II}}\left(p,\boldsymbol{\boldsymbol{\boldsymbol{\omega}}};\mathbf{H}\right)=R_{0}\exp\left(-\frac{c\sigma^{2}}{p\left\Vert \mathbf{H}\boldsymbol{\boldsymbol{\omega}}\right\Vert ^{2}}\right)$ introduced in \cite{vero-TSP-2011},
where $c>0$ is a constant related to the spectral efficiency of the
system and $R_{0}$ the raw transmission rate.
\end{itemize}
A well-admitted transmission \textit{cost function} is as follows \cite{richter-vtc-2009}:
\begin{equation}
C\left(x\right)=C(p, \boldsymbol{\omega}) = p+P_{0}\label{eq:power with circuit power}
\end{equation}
where $P_{0}$ represents a static cost such as the circuit power or the computation power.

Given the fact that the utility consists of the ratio of a concave function to an affine function, it is well known that the resulting utility in Cases I and II is a pseudo-concave (PC) function \cite{zapp-sur-2015}.
 Moreover, another common point between the two above functions is that the beamforming vector only influences the energy-efficiency through the equivalent channel gain $g = \| \mathbf{H}\boldsymbol{\boldsymbol{\omega}} \|^2$ and that the functions are monotonically increasing w.r.t. $g$ for a fixed transmit power. 

 In the presence of a FRF link to acquire the parameter $g$, which corresponds to the channel gain in our problem, the conventional approach consists in quantizing the parameter $g$ and then to report to the transmitter the quantized value of $g$ through the feedback channel. Then, the transmitter maximizes the utility $u$ based on the noisy value of $g$. Note that, with this approach, the decision set may be continuous. With the proposed approach described in Sec. \ref{sec:IWO-DE-algorithm}, the receiver directly reports the decision index to the transmitter and the index rate is chosen to meet the feedback channel requirements such as the finite-rate feedback constraint.
%Motivated by the observation mentioned in the introduction, we thus  assume they now have to lie within finite decision sets and want to determine both the optimal decision sets to be used and the optimal decision to be taken. For this purpose, we resort to the general and initial problem formulation provided in Sec. II. 

Let respectively denote by $M_1$ and $M_2$ the cardinalities of the power level set and the beamforming vector set. These sets are denoted by: $\mathcal{P}=\left\{ p_{1},\dots, p_{M_1}\right\}$ and $\varOmega=\left\{ \boldsymbol{\boldsymbol{\omega}}_{1},\dots,\boldsymbol{\boldsymbol{\omega}}_{M_2}\right\} $. We define the required \textit{amount of feedback information} to take a decision by $B_i = \log_2 M_i$, which expresses in bit per decision. 

In 5G networks, one desirable scenario will be to be able to maximize energy-efficiency under some QoS constraints e.g., for URLLC \cite{nall-GLOBECOM-2018,wang-GLOBECOM-2018}. Obviously, the choice of the transmission decision set can have an impact on the QoS. This is the reason why we should introduce a transmission reliability constraint for the forward communication link (transmitter $\rightarrow$ receiver) and a delay constraint for the reverse of feedback communication link (receiver $\rightarrow$ transmitter). If the data rate from the transmitter to the receiver has to exceed the minimum rate $r_0$, this induces a constraint on the benefit function $V^{i}$. Equally, if  the maximum delay to transfer the channel state information from the receiver to the transmitter is $t_0$, the sum information-rate therefore has to meet the constraint $B_1+B_2 \leq R t_0$, $R$ being the available feedback channel rate.   Having introduced these notations and made these observations, the decision set OP writes in the case of energy-efficient power control and beamforming as:
\begin{align}
\max_{B_{1},B_{2},\mathcal{P},\varOmega} & \ \mathbb{E}_{\mathbf{H}}\left[\frac{V^{i}\left(\widehat{p}_{\mathcal{P}}^{\star}\left(\mathbf{H}\right),\boldsymbol{\widehat{\boldsymbol{\omega}}}_{\varOmega}^{\star}\left(\mathbf{H}\right);\mathbf{H}\right)}{\widehat{p}_{\mathcal{P}}^{\star}\left(\mathbf{H}\right)+P_{0}}\right]\nonumber \\
\text{s.t.} & \ -\mathbb{E}_{\mathbf{H}}\left[V^{\textrm{II}}\left(\widehat{p}_{\mathcal{P}}^{\star}\left(\mathbf{H}\right),\boldsymbol{\widehat{\boldsymbol{\omega}}}_{\varOmega}^{\star}\left(\mathbf{H}\right);\mathbf{H}\right)\right]+r_{0}\leq0\nonumber \\
 & \ B_{1}+B_{2}-R t_{0} \leq0\label{eq:OP for EE}
\end{align}
where 
\begin{equation}
\boldsymbol{\widehat{\boldsymbol{\omega}}}_{\varOmega}^{\star}\left(\mathbf{H}\right)\in\arg\max_{\boldsymbol{\boldsymbol{\boldsymbol{\omega}}}\in\varOmega}\ \left\Vert \mathbf{H}\boldsymbol{\boldsymbol{\omega}}\right\Vert ^{2}\label{eq:quantized optimal beamforming vector}
\end{equation}
and
\begin{equation}
\widehat{p}_{\mathcal{P}}^{\star}\left(\mathbf{H}\right)\in\arg\max_{p\in \mathcal{P}}\ \frac{V^{i}\left(p,\boldsymbol{\widehat{\boldsymbol{\omega}}}_{\varOmega}^{\star}\left(\mathbf{H}\right);\mathbf{H}\right)}{p+P_{0}}.\label{eq:quantized optimal power level}
\end{equation}

\section{Auxiliary results}

The purpose of this section is twofold. First, we show how to simplify the OP under consideration and thus allow it to be solved more easily by the algorithm proposed in Sec. III. Second, we provide the solution of the OP in the limiting case where the number of decisions is infinite, which corresponds to the ideal situation where continuous decisions can be taken (based on perfect CSI feedback typically).

\subsection{Simplification of the original optimization problem}

Solving  OP in (\ref{eq:OP for EE}) with usual numerical techniques may be difficult. The main reasons are as follows:
\begin{enumerate}
\item The OP in (\ref{eq:OP for EE}) is a mixed-integer OP.
\item We are searching the optimal decision set instead of the optimal decision
pair, which is more involving computational speaking. 
\item The quantities $\boldsymbol{\widehat{\boldsymbol{\omega}}}_{\varOmega}^{\star}\left(\mathbf{H}\right)$
and $\widehat{p}_{\mathcal{P}}^{\star}\left(\mathbf{H}\right)$ are defined
in $\arg\max$-form which is difficult to evaluate explicitly.
\end{enumerate}
Therefore, to circumvent the above difficulties we propose to use several key ingredients which allows one to still solve this non-trivial problem.

First, it can be noticed that at optimality all the budget of feedback information should be used that is, at optimality we have that $B_{1}+B_{2}=R t_0$. Second, we can split (\ref{eq:OP for EE}) into some sub-OPs $\textrm{OP}\left(B_{1},B_{2}\right)$
for $\left(B_{1},B_{2}\right)$ fixed. Third, in practice, Monte-Carlo
simulations can be used to approximate the average utility. Indeed, assume
a sequence of $N$ channel realizations $\mathcal{H}=\left\{ \mathbf{H}_{\ell}\right\} _{\ell=1}^{N}$ has been generated, we then only need to solve $\textrm{OP}\left(B_{1},B_{2}\right)$
for every possible configuration $\left(B_{1},B_{2}\right)$. Denote $\Delta[M]$  the set of all subsets with cardinality $M$ of a universal set,
the optimal solution is obviously the configuration that yields the best
performance among all sub-OPs $\textrm{OP}\left(B_{1},B_{2}\right)$, $\left(\mathcal{P},\varOmega\right) \in \Delta[M_1] \times  \Delta[M_2]$, defined as:

\begin{align}
\max_{\left(\mathcal{P},\varOmega\right)} & \ \frac{1}{N}\sum_{\ell=1}^{N}\left\{ \frac{V^{i}\left(\widehat{p}_{\mathcal{P}}^{\star}\left(\mathbf{H}_{\ell}\right),\boldsymbol{\widehat{\boldsymbol{\omega}}}_{\varOmega}^{\star}\left(\mathbf{H}_{\ell}\right);\mathbf{H}_{\ell}\right)}{\widehat{p}_{\mathcal{P}}^{\star}\left(\mathbf{H}_{\ell}\right)+P_{0}}\right\} \nonumber \\
\text{s.t.} & \ \frac{1}{N}\sum_{\ell=1}^{N}\left[V^{\textrm{II}}\left(\widehat{p}_{\mathcal{P}}^{\star}\left(\mathbf{H}_{\ell}\right),\boldsymbol{\widehat{\boldsymbol{\omega}}}_{\varOmega}^{\star}\left(\mathbf{H}_{\ell}\right);\mathbf{H}_{\ell}\right)\right]-r_{0}\geq0.\nonumber \\
\label{eq:OP for EE relaxed for fixed configuration}
\end{align}

\subsection{Solution for the limiting case of continuous decisions}

If the number of feedback bits $B_{1}$ and $B_{2}$ tend to infinity, obviously the OP defined in (\ref{eq:OP for EE}) boils down to a simple continuous OP. The optimal solution of the continuous OP for Utility functions I and II is given by the following proposition \ref{prop:Optimal pair for log}:

\begin{prop}

\label{prop:Optimal pair for log} The pair of beamforming vector
and power level maximizing the Case I and Case II utility function for channel realization $\mathbf{H}$ is given by:
\begin{equation}
\begin{cases}
\boldsymbol{\boldsymbol{\boldsymbol{\omega}}}^{\star}\left(\mathbf{H}\right) & =\boldsymbol{v}\left(\mathbf{H}\right)\\
p_{i}^{\star}\left(\mathbf{H}\right) & =\min\left\{ p_{i}\left(\lambda_{\max}\right),P_{\max}\right\} 
\end{cases}
\end{equation}

where $\boldsymbol{v}\left(\mathbf{H}\right)$ is the dominant
right vector corresponding to the largest singular value of matrix
$\mathbf{H}$ and $\lambda_{\max}=\left\Vert \mathbf{H}\boldsymbol{v}\right\Vert ^{2}$,
$p_{\textrm{I}}\left(\lambda_{\max}\right)$ is the unique solution of the following
equation:
\begin{equation}
\log\left(1+\frac{\lambda_{max}p}{\sigma^{2}}\right)-\frac{p+P_{0}}{p+\frac{\sigma^{2}}{\lambda_{max}}}=0\label{eq:optimal power lever for log}
\end{equation}
and
\begin{equation}
p_{\textrm{II}}\left(\lambda_{max}\right) = 
{\frac{c\sigma^{2}}{2\lambda_{max}}}\left(1+\sqrt{1+\frac{4\lambda_{max}P_{0}}{c\sigma^{2}}}\right)\label{eq:optimal power lever for exp}
\end{equation}

\end{prop}
\begin{IEEEproof}
see Appendix A.
\end{IEEEproof}
It is worth mentioning that the optimal solution in continuous decision set is exactly the one corresponding to the situation where transmitter has perfect CSI (CSIT).

\section{Proposed algorithm \label{sec:IWO-DE-algorithm}}

As already explained, to our knowledge, the problem of determining the optimal decision set has not been tackled in the related wireless literature and the one on energy-efficient communications in particular. Therefore, we had to determine a numerical technique to be able to effectively compute the optimal decision set. A natural way of doing so is to resort to learning. Unfortunately, there doesn't seem to exist a "plug-and-play" learning technique which is directly applicable to our problem. For instance, existing reinforcement learning techniques are not suited to find a decision set in a high dimensional continuous set under acceptable complexity.  A Lloyd-Max-type algorithm could be used for solving OP in (\ref{eq:OP for EE relaxed for fixed configuration}). However it is well known that finding explicit regions for Lloyd-Max-type algorithm is time-consuming. There we propose here one possible way of solving the problem. More advanced techniques may be found or designed, constituting possible extensions of this paper. Here, our goal was to find a suitable candidate which proves the potential of our new approach over the conventional approach (which consists in quantizing the parameter information such as the channel state information).

IWO-DE algorithm is firstly proposed in \cite{cai-SC-2013} by combining IWO \cite{lucas-EI-2006} and DE \cite{storn-JGO-1997} which are essentially two evolutionary algorithms. Evolutionary algorithms have been widely used in many areas with its benefits such as simple computation, robustness and etc (see \cite{fogel-bcec-1997}). We modified the IWO-DE algorithm to adapt our QoS constraint and our search space which is mainly the $\left(2N_{t}-1\right)$-dimensional unit sphere.

It appears that OP in (\ref{eq:OP for EE relaxed for fixed configuration}) is extremely suitable to be
solved by evolutionary algorithms because the  constraint acts as a selection
action, i.e., all generated solutions violate the constraint will
be easily removed. We modified the IWO-DE algorithm to integrate the QoS constraint and deal with our specific search space. To give more focus on the importance of beamforming and for simplicity, we will assume a set of transmit power levels which are equally spaced: $\mathcal{P}=\left\{ 0,\frac{1}{M_1-1}P_{\max},  \frac{2}{M_1-1}P_{\max},  \dots,\frac{M_1-1}{M_1-1}P_{\max}\right\}$.  The fitness function can be naturally constructed from (\ref{eq:OP for EE relaxed for fixed configuration}):
\begin{equation}
U^{i}\left(\mathbf{\varOmega}\right)\coloneqq\frac{1}{N}\sum_{\ell=1}^{N}\left\{ \frac{V^{i}\left(\widehat{p}_{\mathcal{P}}^{\star}\left(\mathbf{H}_{\ell}\right),\boldsymbol{\widehat{\boldsymbol{\omega}}}_{\varOmega}^{\star}\left(\mathbf{H}_{\ell}\right);\mathbf{H}_{\ell}\right)}{\widehat{p}_{\mathcal{P}}^{\star}\left(\mathbf{H}_{\ell}\right)+P_{0}}\right\}. \label{eq:fitness definition}
\end{equation}
To make the computation tasks more convenient, we introduce the matrix  $\mathbf{\boldsymbol{\varOmega}}=\left[\boldsymbol{\boldsymbol{\omega}}_{1},\dots,\boldsymbol{\boldsymbol{\omega}}_{M_{2}}\right]$, which is constructed from the set $\varOmega=\left\{ \boldsymbol{\boldsymbol{\omega}}_{1},\dots,\boldsymbol{\boldsymbol{\omega}}_{M_{2}}\right\} $. The algorithm we propose comprises the following steps:
\begin{itemize}
\item \textbf{Initialization:} randomly choose $W$ beamforming sets in the search
space: $\mathbf{\mathbf{\boldsymbol{\varOmega}}}_{1}^{\left(0\right)},\dots,\mathbf{\boldsymbol{\varOmega}}_{W}^{\left(0\right)}$
as the primitive population. $W$ is called the population size. $\mathbf{\mathbf{\boldsymbol{\varOmega}}}_{k}^{\left(t\right)}$
denotes the $k$-th individual of the $t$-th generation.
\item \textbf{Reproduction}: every individual reproduces its offspring according
to their fitness. The number of offspring for $k$-th
individual at $\left(t+1\right)$-th generation $S_{k}^{\left(t+1\right)}$
is given by:
\begin{equation}
S_{k}^{\left(t+1\right)}=\upsilon\left(\mathbf{\mathbf{\boldsymbol{\varOmega}}}_{k}^{\left(t\right)}\right)\left[S_{M}-S_{m}\right]+S_{m}\label{eq: number of offspring}
\end{equation}
where 
\begin{equation}
\upsilon\left(\mathbf{\mathbf{\boldsymbol{\varOmega}}}_{k}^{\left(t\right)}\right)=\frac{U\left(\mathbf{\mathbf{\boldsymbol{\varOmega}}}_{k}^{\left(t\right)}\right)-\min_{i}U\left(\mathbf{\mathbf{\boldsymbol{\varOmega}}}_{i}^{\left(t\right)}\right)}{\max_{i}U\left(\mathbf{\mathbf{\boldsymbol{\varOmega}}}_{i}^{\left(t\right)}\right)-
\min_{i}U\left(\mathbf{\mathbf{\boldsymbol{\varOmega}}}_{i}^{\left(t\right)}\right)},
\end{equation}
$S_{M}$ and $S_{m}$ are respectively the maximum and minimum numbers of offspring
that an individual is allowed to reproduce.
\end{itemize}

\begin{itemize}
\item \textbf{Spatial Dispersion:} for  $k$-th individual, its offspring obeys a
complex Gaussian distribution $\mathcal{O}_{k}^{\left(t\right)}\sim\mathcal{N}\left(\mathbf{\mathbf{\boldsymbol{\varOmega}}}_{k}^{\left(t\right)},\mu^{\left(t\right)}+\mu^{\left(t\right)}i\right)$.
Then every column vector being essentially a beamforming vector
will be normalized. In what follows, a normalization procedure will
be performed once there is a possibility that new produced beamforming
vector diverges from the unit sphere.  Every individual reproduces its  offspring
in the feasible set till it achieves the number given by  (\ref{eq: number of offspring}).
$\mu^{\left(t\right)}$ is the standard derivation for every entry
of $\mathbf{\boldsymbol{\varOmega}}_{k}^{\left(t\right)}$  controlling
the divergence of the dispersion. The evolution of $\mu^{\left(t\right)}$
through the generations is given by:
\begin{equation}
\mu^{\left(t\right)}=\left(\frac{T-t}{T}\right)^{\gamma}\left[\mu^{ini}-\mu^{end}\right]+\mu^{end}\label{eq:standard derivation evolution}
\end{equation}
\end{itemize}
where $\gamma$ is called the nonlinear index and $\mu^{ini}$ and
$\mu^{end}$ stands for the initial and final standard derivation,
respectively. In general, we should have $\mu^{ini}\gg\mu^{end}$
in order to avoid dropping into a local maximum and $\mu^{end}\rightarrow0$
to increase the accuracy near the potential global optimum.
\begin{itemize}
\item \textbf{Competitive Exclusion:} sort all the offspring together with their
parental individuals in ascending order according to their fitness. Then
select the $W$ first offspring as the original material for next
generation: $\mathbf{\Phi}_{1}^{\left(t\right)},\dots,\mathbf{\Phi}_{W}^{\left(t\right)}$
s.t. $U\left(\mathbf{\Phi}_{1}^{\left(t\right)}\right)\geq\dots \geq U\left(\mathbf{\Phi}_{k}^{\left(t\right)}\right)\geq U\left(\mathbf{\Phi}_{W}^{\left(t\right)}\right)$.
\item \textbf{Mutation:} there are many different
strategies for creating mutations.
For example, for the $k$-th potential individual, we create its possible
mutant by: $\mathbf{\mathbf{\Psi}}_{k}^{\left(t\right)}=\mathbf{\mathbf{\Psi}}_{idx_{1}}^{\left(t\right)}+F_{0}\left(\mathbf{\mathbf{\Psi}}_{idx_{2}}^{\left(t\right)}-\mathbf{\mathbf{\Psi}}_{idx_{3}}^{\left(t\right)}\right)$,
where $F_{0}$ is called the scaling factor.  
And we further choose $idx_{1}=1$ (the best one), $idx_{2}=rand\left(2,W\right)$
and $idx_{3}=rand\left(2,W\right)$ with $idx_{2}\neq idx_{3}$ and
$idx_{2},idx_{3}\neq k$.
\item \textbf{Crossover:} for the $l$-th component of the $k$-th individual at next
generation $\boldsymbol{\omega}_{k,l}^{\left(t+1\right)}$, we let
\[
\boldsymbol{\omega}_{k,l}^{\left(t+1\right)}=\begin{cases}
\boldsymbol{\psi}_{k,l}^{\left(t\right)}, & y_{l}\leq C_{r}\ or\ l=L_{r}\\
\boldsymbol{\phi}{}_{k,l}^{\left(t\right)}, & \textrm{otherwise}
\end{cases}
\]
\end{itemize}
where $\boldsymbol{\omega}_{k,l}^{\left(t+1\right)}, \boldsymbol{\psi}_{k,l}^{\left(t\right)}$
and $\boldsymbol{\phi}{}_{k,l}^{\left(t\right)}$ is the $l$-th component of $\mathbf{\mathbf{\boldsymbol{\varOmega}}}_{k}^{\left(t+1\right)},\mathbf{\mathbf{\Psi}}_{k}^{\left(t\right)}$
and $\mathbf{\Phi}_{k}^{\left(t\right)}$, respectively, $y_{l}$
is a random variable uniformly distributed over $\left[0,1\right]$,
$C_{r}$ is called the crossover probability and $L_{r}$ is a randomly
chosen index so that the mutant decision set can't be identical to
the original one.
\begin{itemize}
\item \textbf{Selection Operation:} only mutant decision increases the fitness ($U\left(\mathbf{\mathbf{\boldsymbol{\varOmega}}}_{k}^{\left(t+1\right)}\right)>U\left(\mathbf{\Phi}_{k}^{\left(t\right)}\right)$)
will be conserved. Otherwise, $\mathbf{\mathbf{\boldsymbol{\varOmega}}}_{k}^{\left(t+1\right)}=\mathbf{\Phi}_{k}^{\left(t\right)}$. 
\end{itemize}
If the initial population is well selected, the population of beamforming
sets will converge to the optimal direction set $\mathbf{\boldsymbol{\varOmega}}^{\star}$
for a sufficiently large number of generations.

\section{Numerical Results\label{sec:Simulation-Results}}

One of the main objectives of this section is to show that the proposed approach may bring some significant improvements when compared to the conventional approach. The conventional approach consists in quantizing the channel state and reporting the corresponding information to the transmitter. In most real systems and existing standards, uniform quantization is implemented. The average available amount of feedback information is thus $\frac{B_{2}}{N_{t}}$ per channel use. Here, we consider a more advanced quantizer namely, the Lloyd-Max (LM) quantizer in \cite{lloyd-TIT-1982}. Essentially, the LM quantizer consists in determining the quantization cells and representatives in an iterative manner to minimize the distortion $\mathbb{E}\left[ \| g - \widehat{g} \|^2 \right]$, $\widehat{g}$ being the quantized channel. This quantized information is then used by the transmitter to maximize its utility function $u^{i}(x; g)$.  We will refer to this algorithm as the \textit{``best conventional approach in SOTA"}. Moreover, the random vector quantization (RVQ) scheme  should be taken as reference as well which is proved to be near-optimal for moderate information feedback of capacity maximization problem in \cite{love-TSP-2007}. For the simulation setting, we will consider a typical scenario defined by: $N_{t}=4$; $N_{r}=1$;  $r_{0}=3\times10^{5}bps$, $t_{0}=0.01s$; \ $R_{0}=10^{6}bps$;  \ $c=0.1$; \ $P_0=0.5 $mW;\ $P_{\max}=1$mW;\ $\sigma^2=1$mW. Similarly, for the the IWO-DE algorithm, a typical setting (in coherence with related evolutionary algorithms) will be assumed as in Table \ref{tab:Parameter-setting-for IWO-DE}. 

\begin{table}[tbh]

\caption{\label{tab:Parameter-setting-for IWO-DE}Parameter setting for IWO-DE}

\centering{}%
\begin{tabular}{|c|c|}
\hline 
Parameters & Value\tabularnewline
\hline 
\hline 
Population size $W$ & 10\tabularnewline
\hline 
number of generations $T$ & 400\tabularnewline
\hline 
max number of offspring $S_{M}$ & 20\tabularnewline
\hline 
min number of offspring $S_{m}$ & 10\tabularnewline
\hline 
Non-linear index $\gamma$ & 2.5\tabularnewline
\hline 
Initial standard derivation $\mu^{ini}$ & $\frac{1}{N_{t}}$\tabularnewline
\hline 
Final standard derivation $\mu^{end}$ & $\frac{1}{200N_{t}}$\tabularnewline
\hline 
Scaling factor $F_{0}$ & 0.9\tabularnewline
\hline 
Crossover probability $C_{r}$ & 0.9\tabularnewline
\hline 
\end{tabular}
\end{table}
In order to clarify the impact of the power and beamforming separately,
we consider two following different situations:
\begin{enumerate}
\item  When the influence of $B_{i}$ is assessed,  $B_{j}$ ($j \neq i$) is fixed.  
\item We fix the total number of quantization bits $B=B_{1}+B_{2}$.
\end{enumerate}

We introduce a key quantity which is the \textit{Relative Optimality Loss:}
\begin{equation}
\sigma \left(\%\right) = \frac{U_{\textrm{CSIT}}  - U_{\textrm{FRF}} }{U_{\textrm{CSIT}}}\times 100\% 
\end{equation}
where $U_{\textrm{CSIT}}$ is the maximum utility achieved by the transmitter when the perfect knowledge of CSI $g$ is available at the transmitter and $U_{\textrm{FRF}}$  is the maximum utility achieved by the transmitter in presence of finite rate feedback (FRF). The latter may assume the conventional approach (relying on the LM quantizer or uniform quantizer) , the proposed approach (relying on the decision set optimization) and RVQ.

\subsection{Separate quantization over beamforming or power levels}
\begin{figure}[htbp]
\centering{}\includegraphics[scale=0.5]{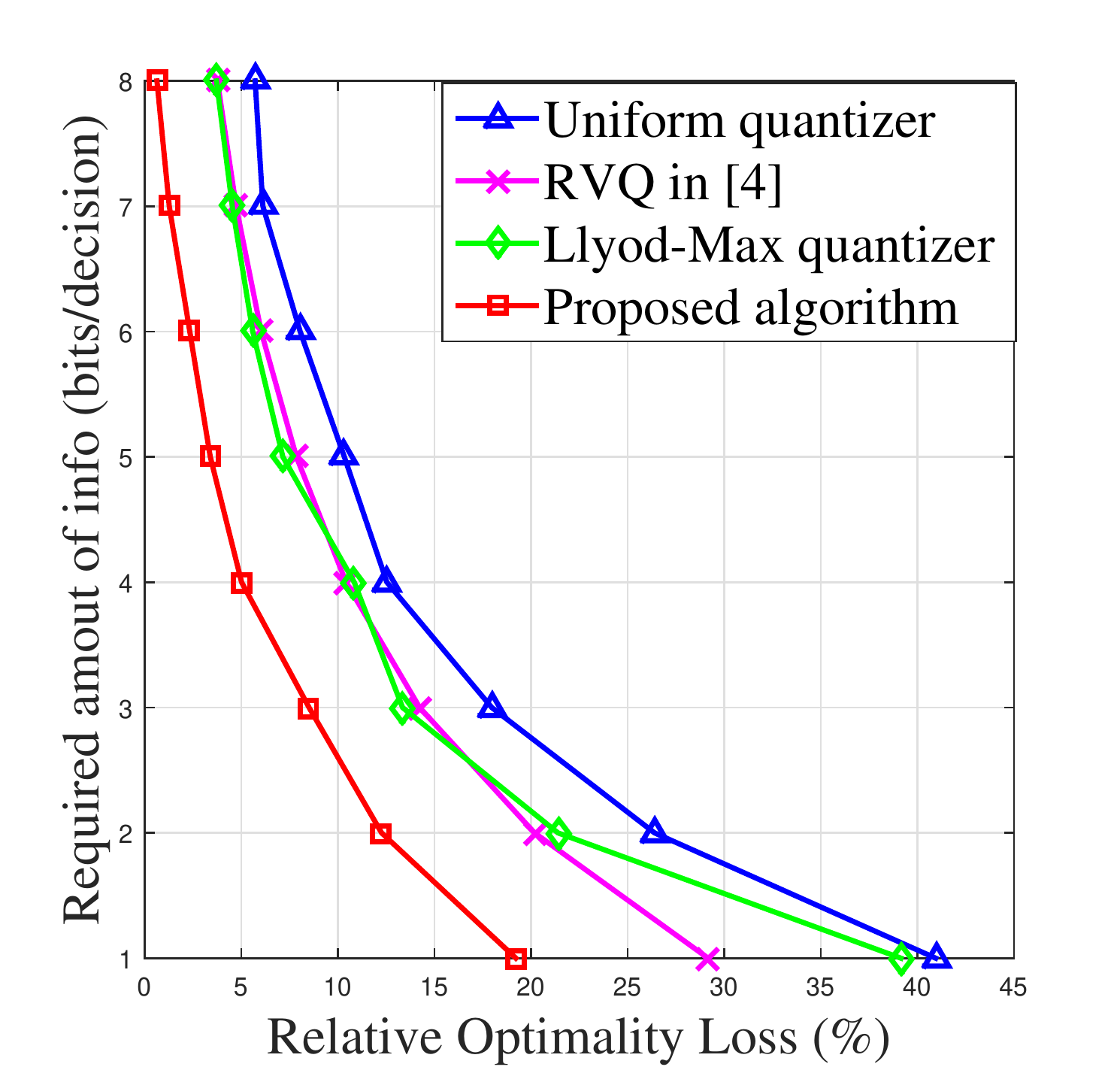}\caption{ The benefits from using our algorithm is very apparent on this figure. For example, for an optimality loss of $5\%$ between the perfect CSI case and the finite-rate feedback case, the amount of information needed to perform beamforming can be reduced by around $2$ by moving from the best state-of-the-art approach to the proposed approach. Here, Case II and $B_{1}=4$ bits/decision are assumed.}\label{fig:compress_rate}
\end{figure}

\begin{figure}[htbp]
\centering{}\includegraphics[scale=0.45]{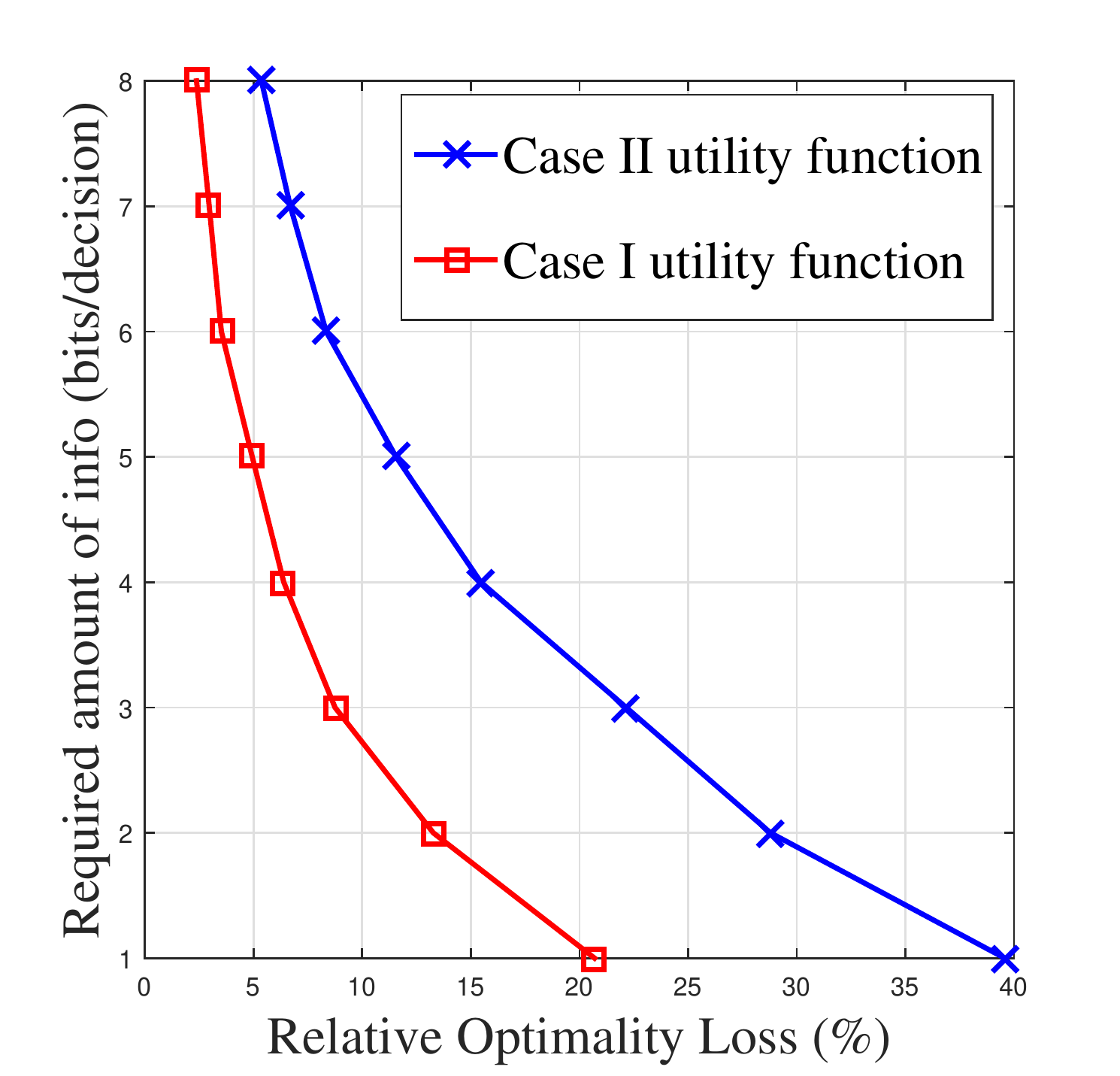}\caption{Here, the influence of the utility function on the obtained results is assessed. It is clearly seen that some utilities require more feedback than others. Here, it is seen that considering the packet success rate for the benefit function (Case II) requires more feedback resources than using the capacity function (Case I). Remarkably, it is possible to quantify this extra amount of resources. Here, it is assumed that $B_{1}=4$ bits/decision.\label{fig:compress_rate_B1_varing}}
\end{figure}

First of all, we fix $B_{1}=4$ bits and analyze the influence of $B_{2}$ to the relative optimality loss. To compare our approach with the conventional approach, Fig. \ref{fig:compress_rate} illustrates the required amount of information for beamforming quantization  to achieve a given relative optimality loss of case I utility function. For a given same optimality loss, remarkably, one can observe that with our approach one can \emph{reduce by a factor 2 the amount of beamforming bits} with respect to the LM quantizer and RVQ. In addition, if the number of  bits allocated to beamforming quantization is quite small,  the relative optimality loss remains acceptable for the proposed approach while it is large for other existing solutions. Moreover, to explore the impact of utility function on beamforming quantization. Fig. \ref{fig:compress_rate_B1_varing} compares the required $B_{2}$ to achieve a given optimality loss between the utility functions in case I and the utility function in case II. By implementing the proposed quantization scheme, the minimum number  of bits for EE of case II is larger than the EE of case I  for achieving the same performance which may suggests that  the EE of case II (packet transmission success rate) is slightly sensitive to the quality of quantization
than EE of case I (capacity) and thus worth more feedback bits and better beamforming code book design techniques. 

\begin{figure}[htbp]
\centering{}\includegraphics[scale=0.48]{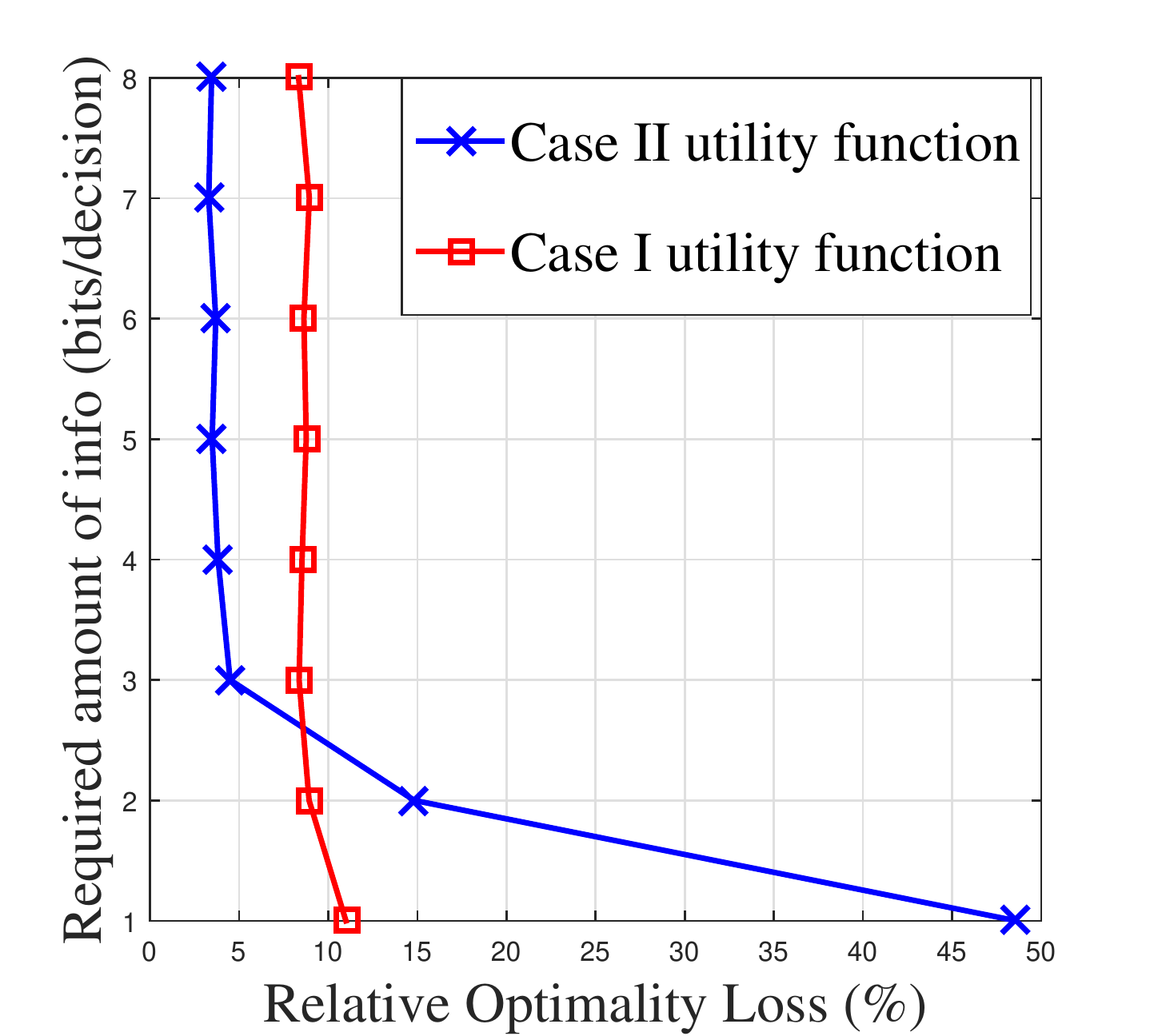}\caption{The impact of $B_2$ on the preceding observation is assessed. Here, with $B_{2}=6$ bits/decision, the curves are much steeper, indicating that the choice of the number of feedback rate is more critical in this regime as soon as small optimality losses are desired.\label{fig:optimality_loss_B2_varying}}
\end{figure}

To see the influence of the power level quantization, we fix the bits of beamforming quantization as $B_{2}=6$
and vary the bits for power level feedback from $1$ to $8$.  Fig. \ref{fig:optimality_loss_B2_varying} shows the evolution of  required amount of information for power quantization as  function of relative optimality loss. For EE of case I , different from EE of case II, increasing
the number of bits for power quantization have less important impact on performance of the system. This improvement is always modest while the improvement is firstly sharp when few bits are available but then becomes modest with enough number of bits provided for EE of case II.
Thus if the bits for beamforming quantization are sufficient even one bit feedback information about power provides acceptable performance for EE of case I. Up to now, We can conclude that EE of case II  is sensitive to both the quality of beamforming quantization and power quantization combing the observation in Fig. \ref{fig:compress_rate}
and Fig. \ref{fig:compress_rate_B1_varing}. We need to further find the optimal  bits allocation policy for EE of case II in Section \ref{joint_quantization}.


\subsection{Joint Quantization\label{joint_quantization}}

\begin{figure}[htbp]
\centering{}\includegraphics[scale=0.5]{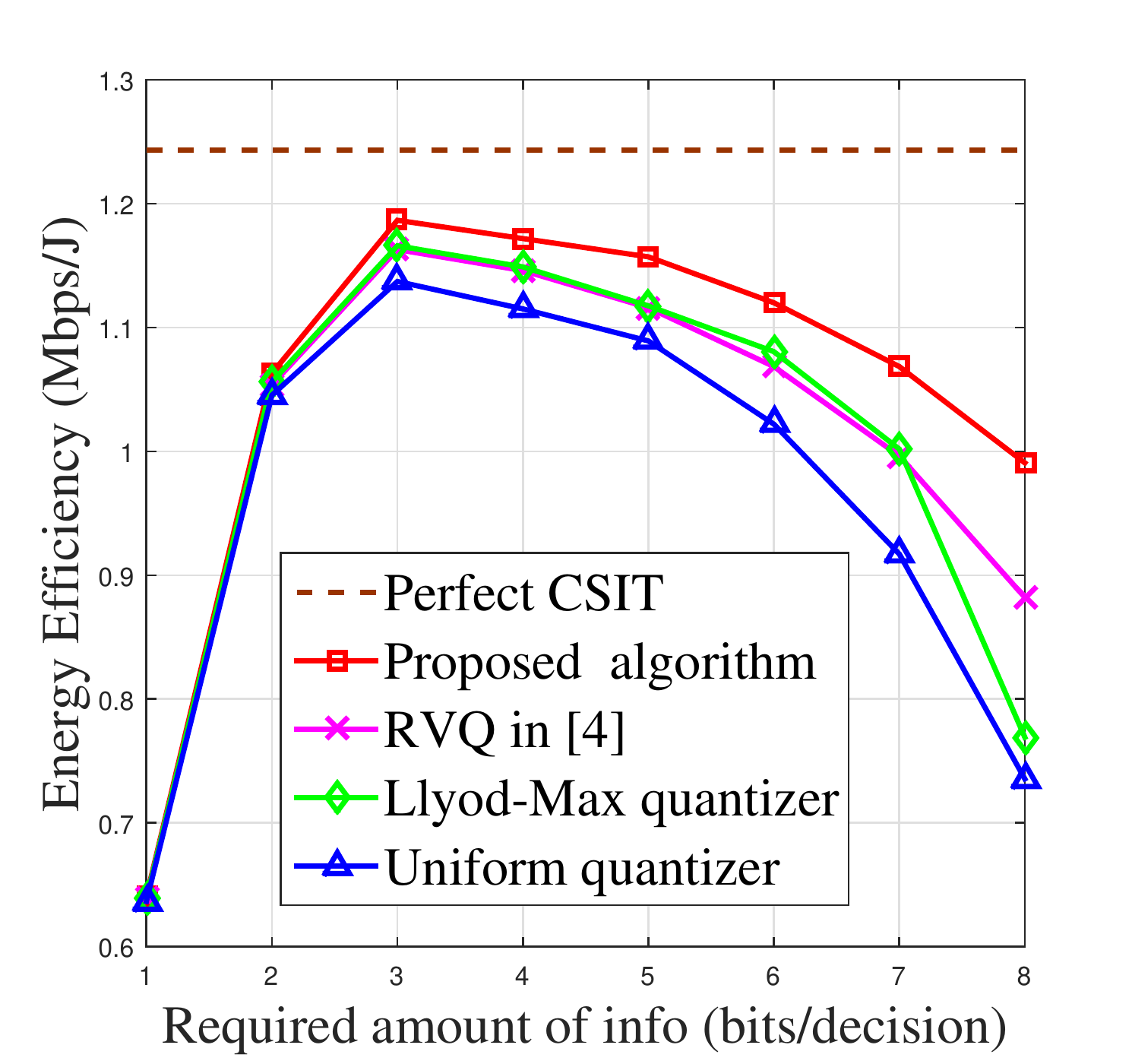}\caption{Here, the gain brought by our approach are measured in terms of EE for case II.  The conventional approach is sensitive to the available amount of feedback information for beamforming when power level quantization is rough while the proposed approach offers good performance for a large range of feedback rates.
\label{fig:Energy-Efficiency B fixed}}
\end{figure}

According to the precedent observations, finding a proper allocation policy between $B_{1}$ and $B_{2}$ is necessary. In order to determine the optimal allocation  of bits for EE of case II , we assume that the total quantization bits are
fixed so that the transmitter merely seeds the essential information back to  the receiver.
We fix the total number of bits for quantization as $B=8$ (exactly one byte). Fig.
\ref{fig:Energy-Efficiency B fixed} shows the evolution of energy
efficiency of case II  as function
of quantization bits used for beamforming. To achieve the best performance,
among $8$ total quantization bits, we should allocate  $3$
bits for beamforming quantization and $5$ bits for power quantization. Moreover, for all methods, sufficient number of bits should be conserved to beamforming quantization by observing the sharp decay of average utility for $1\leq B_{2} \leq 3$.  Finally, even no information provided for power level ($B_{2}=8$), the energy efficiency achieved by our proposed approach  and RVQ is acceptable which shows the importance of quantizing directly on the decision itself instead of quantizing the CSI in the conventional approach.

\vspace{-0.1in}

\section{Conclusions\label{sec:Conclusions}}

This paper treats for the first time the problem of finding jointly set of power level and beamforming vectors for energy-efficient communications. While the problem is relatively easy to solve when decisions can be continuous, the problem needs to be formulated properly when the decision set is imposed to be finite. We precisely propose this new formulation of the problem. The problem of determining an optimal decision set appears. Whereas the latter problem may be difficult in general, we show that it can be solved for the wireless application of interest. For this, we propose a new algorithm, which is referred to the IWO-DE algorithm. It allows us to search for optimal sets of beamforming vectors over the ($2N_{t}-1$)-dimensional unit sphere. When applied to energy-efficient communications, our approach is shown to outperform the best state-of-the-art techniques such as Lloyd-Max algorithm and RVQ.  Obviously, our approach needs to be explored and developed further. In particular, when the system dimensions increase, complexity issues need to be considered. When there is interference, the proposed framework needs to be extended. In the presence of interactions between the decision-makers, other issues such as Braess paradoxes may arise and make the problem even more challenging. 

\section*{Acknowledgment}

This work has been funded by the RTE-CentraleSupelec Chair on ``The Digital Transformation of Electricity Networks".

\section*{Appendix A\label{sec:Appendix-A}}

For a fixed power $p$, maximizing the utility I is equivalent to maximize
the equivalent channel gain $ g =\left\Vert \mathbf{H}\boldsymbol{w}\right\Vert ^{2}$.
Thus the optimal beamforming vector is given by the right dominant vector
$\boldsymbol{v}$ which is the right vector corresponding to the
largest singular value $\sqrt{\lambda_{max}}$ of $\mathbf{H}$.

Thus the maximum of equivalent channel gain is given by $G_{max}=\left\Vert \mathbf{H}\boldsymbol{v}\right\Vert ^{2}=\lambda_{max}$, then 
\begin{align*}
\frac{\partial u}{\partial p} & =\frac{-\ln\left(1+\frac{\lambda_{max}p}{\sigma^{2}}\right)+\frac{p+P_{0}}{p+\frac{\sigma^{2}}{\lambda_{max}}}}{\left(p+P_{0}\right)^{2}}
\end{align*}
Define $f\left(p\right)=-\ln\left(1+\frac{\lambda_{max}p}{\sigma^{2}}\right)+\frac{p+P_{0}}{p+\frac{\sigma^{2}}{\lambda_{max}}}$,
one has:
\[
f^{'}\left(p\right)=-\frac{1}{p+\frac{\sigma^{2}}{\lambda_{max}}}-\frac{\frac{\sigma^{2}}{\lambda_{max}}-P_{0}}{\left(p+\frac{\sigma^{2}}{\lambda_{max}}\right)^{2}}
\]
\begin{itemize}
\item If $\frac{\sigma^{2}}{\lambda_{max}}\geq P_{0}$ then $f^{'}\left(p\right)<0$
for $\forall p>0$ meaning $f\left(p\right)$ is a monotonically decreasing
function. Furthermore, we have $f\left(0\right)=\frac{\lambda_{max}P_{0}}{\sigma^{2}}>0$
and $f\left(+\infty\right)=-\infty<0$, there exists an unique solution
$p_{\textrm{I}}\left(\lambda_{max}\right)$ of the Eq. \ref{eq:optimal power lever for exp}
by mean-value theorem.
\item If $\frac{\sigma^{2}}{\lambda_{max}}<P_{0}$ then $f^{'}\left(p\right)>0$
for $0<p<P_{0}-\frac{\sigma^{2}}{\lambda_{max}}$ and $f^{'}\left(p\right)>0$
for $0<p<P_{0}-\frac{\sigma^{2}}{\lambda_{max}}$. Therefore $\max f\left(p\right)=f\left(P_{0}-\frac{\sigma^{2}}{\lambda_{max}}\right)>f\left(0\right)>0$.
With the same argument, there exists an unique solution $p_{\textrm{I}}\left(\lambda_{max}\right)$
of the Eq. \ref{eq:optimal power lever for exp} in $\left(P_{0}-\frac{\sigma^{2}}{\lambda_{max}},+\infty\right)$.
\end{itemize}
Moreover we have $f\left(p\right)>0$ for $0<p<p_{\textrm{I}}\left(\lambda_{max}\right)$
and $f\left(p\right)<0$ for $p>p_{\textrm{I}}\left(\lambda_{max}\right)$. Finally
the optimal power $p^{\star}\left(\mathbf{H}\right)$ for channel
realization $\mathbf{H}$ is given by:
\[
p^{\star}\left(\mathbf{H}\right)=\min\left\{ p_{\textrm{I}}\left(\lambda_{max}\right),P_{max}\right\} 
\]
The proof for Case II utility function can be obtained by using similar arguments above.

\end{document}